\documentclass[%
 reprint,
 superscriptaddress,
 amsmath,amssymb,
 aps,
 pra,
]{revtex4-2}

\usepackage[utf8]{inputenc}
\usepackage[T1]{fontenc}
\usepackage{textcomp}       
\usepackage{eurosym}        
\usepackage{graphicx}
\usepackage{dcolumn}
\usepackage{bm}
\usepackage{upgreek}
\usepackage{xcolor}
\usepackage{newunicodechar}
\newunicodechar{ }{\,}
\usepackage{hyperref}
\hypersetup{hidelinks}

\begin{document}
\title{Octave-spanning, deterministic single soliton generation in 4H-silicon carbide-on-insulator microring resonators}

\author{Yi Zheng}
\thanks{These authors contributed equally.}
\affiliation{DTU Electro, Department of Electrical and Photonic Engineering, Technical University of Denmark, 2800 Kongens Lyngby, Denmark}

\author{Liping Zhou}
\thanks{These authors contributed equally.}
\affiliation {State Key Laboratory of Functional Materials for Informatics, Shanghai Institute of Microsystem and Information Technology, Chinese Academy of Sciences, Shanghai, 200050, China}

\author{Chengli Wang}
\thanks{These authors contributed equally.}
\affiliation{State Key Laboratory of Functional Materials for Informatics, Shanghai Institute of Microsystem and Information Technology, Chinese Academy of Sciences, Shanghai, 200050, China}

\author{Yanjing Zhao}
\affiliation{DTU Electro, Department of Electrical and Photonic Engineering, Technical University of Denmark, 2800 Kongens Lyngby, Denmark}

\author{Ailun Yi}
\affiliation {State Key Laboratory of Functional Materials for Informatics, Shanghai Institute of Microsystem and Information Technology, Chinese Academy of Sciences, Shanghai, 200050, China}

\author{Kresten Yvind}
\affiliation{DTU Electro, Department of Electrical and Photonic Engineering, Technical University of Denmark, 2800 Kongens Lyngby, Denmark}

\author{Xin Ou}
\thanks{ouxin@mail.sim.ac.cn}
\affiliation{State Key Laboratory of Functional Materials for Informatics, Shanghai Institute of Microsystem and Information Technology, Chinese Academy of Sciences, Shanghai, 200050, China}

\author{Minhao Pu}
\thanks{mipu@dtu.dk}
\affiliation{DTU Electro, Department of Electrical and Photonic Engineering, Technical University of Denmark, 2800 Kongens Lyngby, Denmark}

\date{\today}


\begin{abstract}
The miniaturization of self-referencing frequency comb systems enables emerging applications in metrology and spectroscopy. One major challenge in realizing the chip-scale self-referencing function is to generate octave-spanning soliton microcombs with low operation power. Accessing soliton states is also not trivial due to the thermal effect. Though an auxiliary laser was utilized to compensate for the thermal effect, deterministic single soliton generation is still elusive, especially for broadband operation. In this work, dispersion management is performed for a 4H-silicon carbide-on-insulator (SiCOI) multi-mode microring resonator, benefiting from the submicron-confinement waveguide layout. The fundamental transverse electric (TE) mode is engineered to anomalous dispersion for two dispersive waves generation over an octave span. While a higher order TE mode is engineered to normal dispersion to accommodate the auxiliary light for thermal compensation. The normal dispersion prevents modulation-instability Kerr comb generation, allowing for a large soliton existence range. We achieve microring resonators with Q up to 5.8 million and sub-milli-watt-threshold Kerr comb generation. Combining the dispersion-managed design and high Q device, we demonstrate the deterministic generation of a single soliton comb spanning beyond an octave with a low on-chip power of 60 mW. Our demonstration paves the way to realize chip-scale, turn-key, self-referenced frequency combs.
\end{abstract}

\maketitle


\section{\label{sec:intro}Introduction}

Chip-based optical frequency comb has been of great interest in the past decades because it has enormous potential for various applications, from precision metrology to optical synthesis and astronomical spectrograph calibration \cite{Kippenberg2018DissipativeMicroresonators}. Many of these applications require stabilized combs as they provide absolute frequency references. An octave-spanning comb enables f-2f self-referencing, which is essential for comb stabilization \cite{Hansch2006Nobel, Hall2006NobelFrequencies, Spencer2018AnPhotonics}. General methods to obtain a broadband frequency comb on a chip include waveguide-based supercontinuum generation (SCG) and microresonator-based Kerr comb generation (KCG). Octave-spanning SCG typically relies on femtosecond pulses with high peak power, which is beyond the reach of on-chip mode-locked lasers \cite{Gaeta2019Photonic-chip-basedCombs}. Recent studies show that the electro-optic modulation (EOM) of a laser can also be used as a pump source for SCG, but it is still challenging to fulfill the pulse requirement for octave-spanning SCG \cite{Carlson2018UltrafastControl}. An elegant approach is a microresonator-based optical frequency comb (“microcomb” hereafter) generation with a single-frequency pump \cite{DelHaye2007, Herr2014TemporalMicroresonators}. The recent technological development on KCG has shown the potential of a fully integrated microcomb source \cite{Spencer2018AnPhotonics, Stern2018Battery-operatedGenerator, Shen2020IntegratedMicrocombs, Xiang2021LaserSilicon}. However, a fully integrated octave-spanning microcomb that holds the promise of the aforementioned applications has not been realized yet. Up until now, only limited studies have reported octave-span microcomb generation in integrated platforms, including silicon nitride (Si$_3$N$_4$) \cite{Spencer2018AnPhotonics, Li2017StablyRegime, Pfeiffer2017Octave-spanningMicroresonators, Briles2021HybridComb} aluminum nitride (AlN) \cite{Weng2021DirectlyMicroresonator, Liu2021AluminumSelf-referencing}, and lithium niobate (LN) \cite{Gong2020Near-octaveMicrocomb, He2021Octave-spanningMicrocombs}. Compared with a narrow-band microcomb, high pump power, typically larger than 100 mW, is required to achieve a microcomb with an octave-spanning frequency range, which makes it challenging to access such a soliton comb in a deterministic manner because of the large thermal effect induced by considerable intracavity power. 

\begin{figure*}
\centering\includegraphics[width=15.5cm]{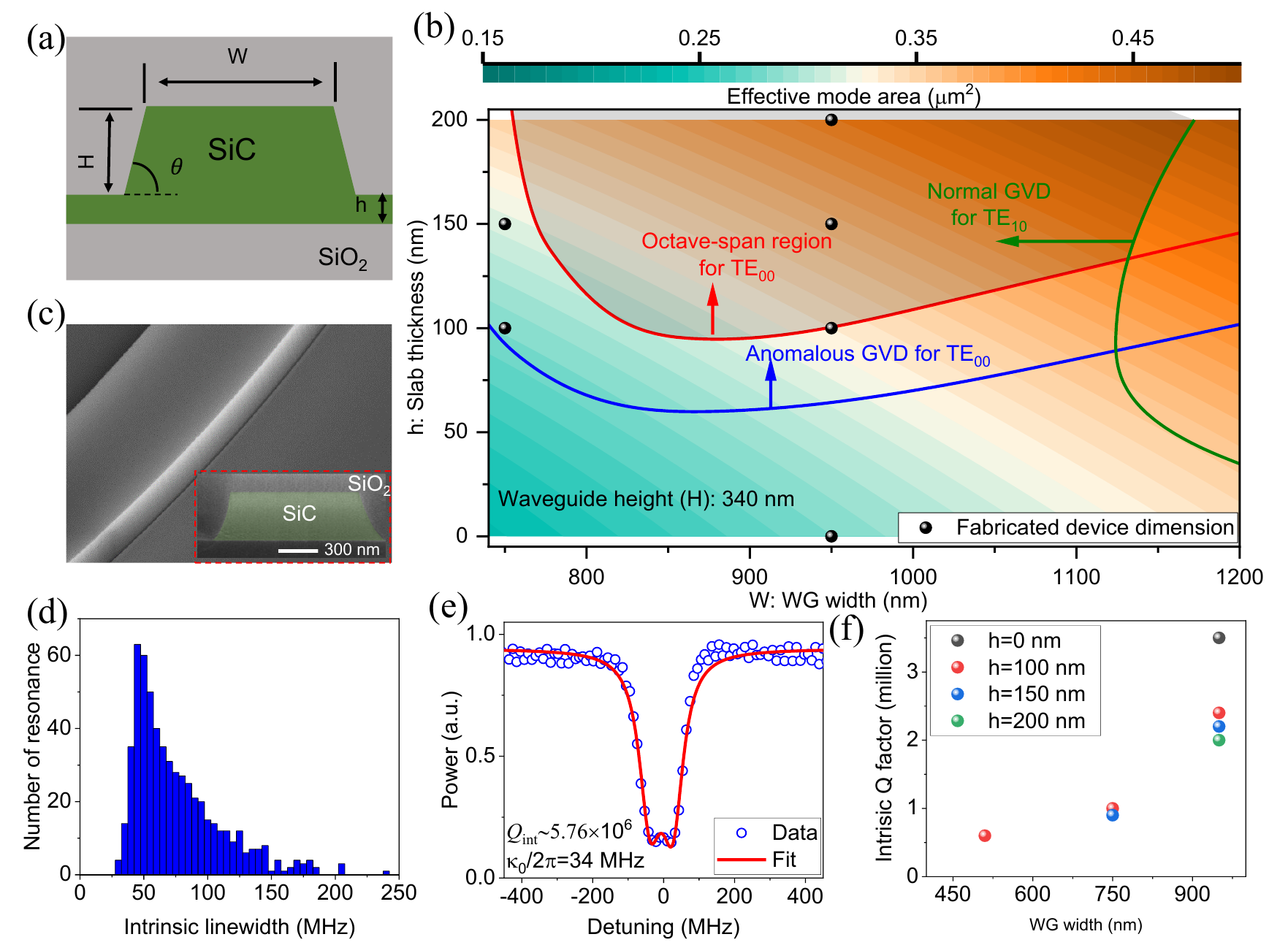}
\caption{\label{fig:design}\textbf{Design and characterization of SiCOI microring resonators.} (a) Schematic cross-section of a SiCOI waveguide defined by waveguide height (${\rm H}$), waveguide width (${\rm W}$), slab thickness (${\rm h}$), and sidewall angle ${\rm \theta}$. (b) Calculated effective mode area (color-shaded contour) and the group velocity dispersion (GVD) (solid-line contour) for the fundamental ${\rm TE_{00}}$ mode (the comb operation mode) of SiCOI waveguides with different cross-section dimensions at 1560~nm. The green contour line represents the zero GVD design for the ${\rm TE_{10}}$ mode (the laser cooling mode). The grey-shaded area is the target design region. (c) Scanning electron microscopy (SEM) image of a fully etched microring resonator waveguide with its cross-sectional view shown in the inset. (d) Histogram of intrinsic resonance linewidth for ten fully-etched microring resonators with the same waveguide dimension (${\rm H = 340}$~nm, ${\rm W = 950}$~nm, ${\rm h = 0}$~nm, and ${\rm \theta = 75^{\circ}}$), showing the most probable value of 55~MHz. (e) Measured (normalized) transmission spectrum for a split resonance showing the highest ${\rm Q}$ of 5.76~million (intrinsic linewidth ${\rm \gamma_{int}}$ of 34~MHz). (f) Intrinsic ${\rm Q}$ value for microring resonators with different waveguide widths (${\rm W}$) and slab thicknesses (${\rm h}$), here ${\rm H = 340}$~nm and ${\rm \theta = 75^{\circ}}$.}
\end{figure*}

Several investigations have been done to overcome the difficulty. For instance, dispersion and cavity-length design to mitigate the issue in a single-cavity, single-pump system, but the method is highly dependent on the free spectral range (FSR) and has so far been demonstrated only for microcombs with THz repetition rates \cite{Wu2023AlGaAsTemperature, Jacobsen2025High-powerGeneration}. Moreover,  fast pump sweeping and single-sideband suppressed-carrier methods with a scan speed up to 100 GHz/$\mu$s were applied to obtain the soliton state \cite{Guo2017UniversalMicroresonators, Stone2018ThermalCombs}. Power-kicking was also successfully implemented for single soliton generation \cite{Yi2016}. Despite those effective methods to mitigate the influence of the thermal effects in the resonator, it is still challenging to implement them in material platforms with large thermal-optic coefficients, including silicon carbide \cite{Guidry2022QuantumMicrocombs, Wang2021High-QPhotonics,Cai2022Octave-spanningPlatform}, gallium nitride \cite{Zheng2022IntegratedPhotonics}, and aluminum gallium arsenide \cite{Zheng2018, Moille2020DissipativeMicroresonator,Ye2024MultimodePhotonics, Chen2024IntegratedMicrocomb} for octave-spanning soliton generation. One can also use a single laser to pump coupled spatial modes in a microresonator, with one mode for comb generation and the other for thermal compensation \cite{Li2017StablyRegime, Weng2022Dual-modeSolitons}. However, this scheme requires delicate control of the two spatial modes, and undesired mode coupling between them may prevent soliton generation \cite{Herr2014ModeMicroresonators}. Alternatively, counter-propagating auxiliary laser light in a microresonator can assist soliton comb generation by mitigating the thermal-optic effect of the comb operation mode \cite{Zhou2019SolitonMicrocavities, Zhao2025ThermalGeneration}. This so-called laser cooling technique not only enables soliton generation but also significantly reduces the thermal noise of the resonator \cite{Drake2020ThermalSolitons}. Nevertheless, previous studies often use the same spatial mode for comb operation and cooling \cite{Wang2018Ultrahigh-efficiencyWaveguides, Zhou2019SolitonMicrocavities, Lu2019DeterministicMicro-resonator, Wang2022SolitonPlatform}. Consequently, the cooling effect may be limited in high-power pumping cases because modulation instability (MI) combs will be generated due to the cooling laser, leading to additional noise in the system, especially for octave-spanning soliton generation where high pump power is required \cite{Pfeiffer2017Octave-spanningMicroresonators}. Furthermore, the self-injection locking technique can bypass the thermal issues and lead to turnkey soliton operation \cite{Shen2020IntegratedMicrocombs, Briles2021HybridComb}. To date, it is still challenging to deterministically generate octave-span single solitons.

In this work, we explore deterministic octave-span single soliton comb generation in the silicon carbide-on insulator (SiCOI) platform with a relatively large thermal-optic coefficient by utilizing the laser cooling technique in dispersion-managed multi-mode microring resonators. We demonstrate high Q microring resonators with sub-micron waveguide cross-sectional dimensions. We achieve an average Q of 3.5 million and the highest Q of 5.8 million in high-confinement microring resonators. The obtained high Q and optimized effective nonlinearity reduce the threshold power for soliton generation, which is very important for laser integration. Moreover, we propose a method for deterministic octave-span single soliton generation using the laser cooling technique in a dispersion-managed multi-mode microring resonator, where the mode for comb pumping keeps low anomalous dispersion while the mode for laser cooling is normal dispersion. For the first time, we accomplish broadband single soliton microcombs in SiCOI microring resonators with low on-chip power of 60  mW, where its bandwidth covers a spectral range beyond one octave (136-307 THz) along with two separated dispersive waves (DWs). With the new laser cooling scheme, we can deterministically generate octave-spanning soliton with a long soliton step of 5 GHz, orders of magnitude longer than previous results \cite{Guidry2022QuantumMicrocombs, Wang2022SolitonPlatform}. Moreover, the microcomb shows improved noise properties at the output. The proposed scheme applies to other material platforms with large thermo-optic coefficients for broadband soliton comb generation. The combination of low-power operation and deterministic octave-spanning, single soliton generation shows the great potential of the SiCOI platform for chip-scale octave-spanning soliton microcombs.

\section{\label{sec:principle}Methods and experimental results}

\subsection{Device design}

SiC has emerged as a promising material for nonlinear photonics \cite{Zheng2019High-qualityCarbide-on-insulator, Yi2020Wafer-scaleDevices, Wang2022SolitonPlatform, Song2019Ultrahigh-QCarbide}. In addition to high nonlinear refractive index (${\rm n_2}$ \cite{DeLeonardis2017DispersionCarbide}), it offers decent ${\rm \chi^{(2)}}$ coefficient (6--18~pm/V \cite{Sato2009AccurateCarbide, Niedermeier1999Second-harmonicPolytypes, Lundquist1995}). Leveraging the large ${\rm \chi^{(2)}}$ nonlinearity, efficient second-harmonic generation (SHG) has been reported in SiC waveguides \cite{Zheng2025EfficientNanowaveguides} and microring resonators \cite{Lukin20204H-silicon-carbide-on-insulatorPhotonics}. Even though efforts have been made to realize octave-spanning combs in SiC \cite{Cai2022Octave-spanningPlatform}, they have not been coherent and thus impossible to self-reference. Only narrow-band ($<$150~nm) soliton microcombs were demonstrated \cite{Guidry2022QuantumMicrocombs, Wang2022SolitonPlatform, Li2024SiliconCommunications}. The main challenge comes from the large thermal nonlinearities in high-quality factor (${\rm Q}$) resonators. Therefore, only a short soliton existence range (SER) of 0.014~GHz was obtained even in a cryogenic environment with only a coupled pump power of 2.3~mW \cite{Guidry2022QuantumMicrocombs}. Though the laser cooling technique has shown its effectiveness in SiCOI microdisk resonators for soliton generation with an on-chip pump power of 300~mW, the single soliton generation is not deterministic, and the comb bandwidth is narrow due to a limited degree of freedom in the dispersion engineering of the microdisk resonators \cite{Wang2022SolitonPlatform}. 

Dispersion engineering is essential for broadband comb generation, especially for octave-spanning comb via dispersive wave generation. For such broadband comb generation, the required high intracavity power usually associates with a strong thermal effect that prohibits the deterministic generation of single solitons even when a counter-propagating auxiliary light is utilized in a microresonator to optically compensate for the thermal effect. Since the auxiliary laser (or the cooling laser) is typically set to a similar power level as the pump laser, an unstable MI comb can be generated from the auxiliary light in the counter-propagating direction, and any unintentional reflection of this auxiliary comb may also degrade the noise property of the soliton output. Therefore, stable and noise-free laser cooling can be expected if the MI comb can be suppressed by using a normal-dispersion waveguide mode to accommodate the auxiliary light. This can be done by assigning the pump and cooling light to two spatial waveguide modes with different dispersion properties. This requires proper dispersion management for a multi-mode microresonator. 

To make efficient dispersion engineering, we investigate the high-confinement SiCOI waveguide with its cross-section shown in Fig.~\ref{fig:design}(a), which is defined by waveguide height (${\rm H}$), waveguide width (${\rm W}$), sidewall angle (${\rm \theta}$), and slab thickness (${\rm h}$). The strong light confinement also results in a small effective mode area and enhances the nonlinear light-matter interaction, which lowers the threshold for KCG. Figure~\ref{fig:design}(b) shows the calculated effective mode area (color-shaded contour) for the fundamental TE mode (${\rm TE_{00}}$ mode) of the SiCOI waveguides with different waveguide widths (${\rm W}$) and slab thicknesses (${\rm h}$). Here the waveguide height (${\rm H}$) is fixed at 340~nm. It is seen in Fig.~\ref{fig:design}(b) that the SiCOI waveguide exhibits a sub-micron effective mode area which varies from 0.15~\textmu m$^2$ to 0.5~\textmu m$^2$ for different dimensions. We also calculated the group velocity dispersion (GVD) for both the ${\rm TE_{00}}$ and ${\rm TE_{10}}$ modes at the wavelength of 1560~nm. The blue and green curves correspond to the zero-GVD waveguide dimensions for the ${\rm TE_{00}}$ and ${\rm TE_{10}}$ modes, respectively. The red contour line corresponds to the waveguide designs which support the generation of two dispersive waves with just an octave frequency separation, while the grey-shaded region shows the waveguide designs with the dispersive wave frequency separation larger than an octave. Here, the ${\rm TE_{00}}$ and ${\rm TE_{10}}$ modes are selected as pump and cooling modes, respectively. And one can use a waveguide design in the grey-shaded region shown in Fig.~\ref{fig:design}(b) for deterministic soliton generation, where the comb operation mode has anomalous dispersion for octave-spanning comb generation while the laser cooling mode exhibits normal dispersion for Kerr comb suppression.

\begin{figure*}[t]
\includegraphics[width=15.5cm]{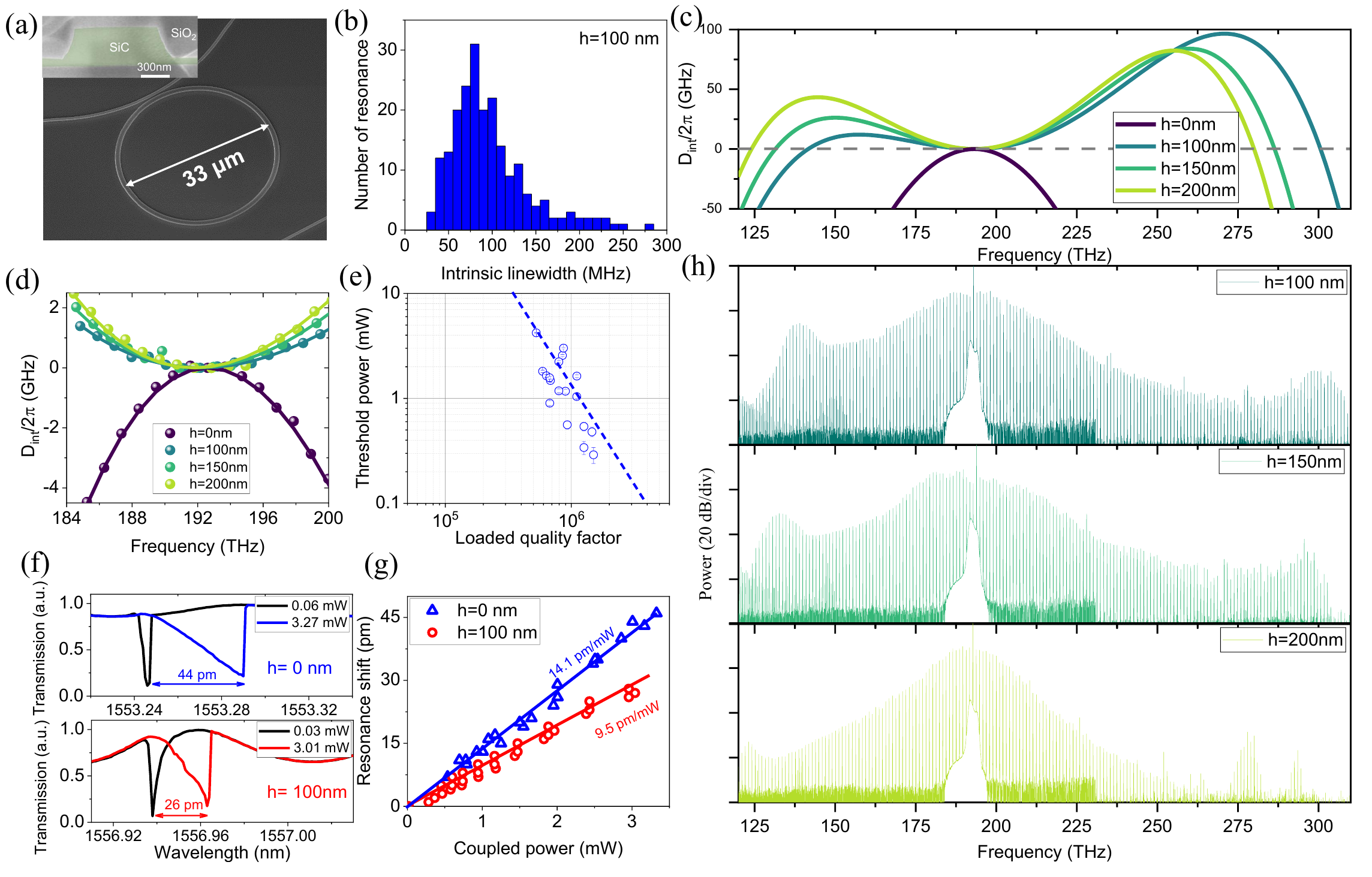}
\caption{\label{fig:engineering} \textbf{Dispersion engineering for SiCOI microring resonators and octave-spanning Kerr comb generation.} (a) An SEM image of a 1-THz-FSR microring resonator with its waveguide cross-sectional view is shown in the inset. (b) Histogram of intrinsic resonance linewidths of ten 1-THz-FSR microring resonators with the waveguide dimension (${\rm H = 340}$~nm, ${\rm W = 950}$~nm, ${\rm h = 100}$~nm, and ${\rm \theta = 75^{\circ}}$). (c) Simulated integrated dispersion of the ${\rm TE_{00}}$ mode for 16.5-\textmu m-radius bent waveguides with different waveguide slab thicknesses ${\rm h}$. (d) Measured integrated dispersion of microring resonators at a frequency range from 184--200~THz with the waveguide designs shown in (c). (e) Measured threshold power of optical parametric oscillation versus ${\rm Q}$ for same dimension devices with the lowest value of 0.29~mW. The error bars correspond to the uncertainty in estimating fiber-to-chip coupling efficiency. The dashed line corresponds to the estimated threshold power at the critical-coupling condition. (f) Measured (normalized) transmission spectra showing thermal resonance shifts for 1-THz-FSR microring resonators with and without a slab. (g) Measured resonance shifts as a function of coupled power for devices in (f). For both waveguide designs, data points are extracted from the characterization of three microring resonators with similar ${\rm Q}$s and coupling conditions. (h) Measured spectra for the octave-spanning Kerr comb generation in devices with different slab thicknesses in (c).}
\end{figure*}

A high ${\rm Q}$ microring resonator with a small effective mode area is desired for low threshold comb generation. We fabricated devices with different waveguide cross-sectional dimensions to investigate their ${\rm Q}$ performance. The black dots in Fig.~\ref{fig:design}(a) indicate the waveguide dimension designs under investigation. The detailed device fabrication procedures can be found in Supporting information. Figure~\ref{fig:design}(c) shows the scanning electron microscopy (SEM) picture of the smooth sidewall of a fully etched microring resonator. The inset is the cross-section of a fully etched waveguide with a dimension of ${\rm W \times H}$. We characterize different microring resonators with the same dimension to obtain the most probable ${\rm Q}$ value for a particular waveguide dimension. Figure~\ref{fig:design}(d) presents the histogram of intrinsic linewidth from ten fully etched devices with a FSR of 109~GHz and shows a most probable value of 55~MHz, corresponding to an intrinsic ${\rm Q}$ factor of 3.5 million. The highest ${\rm Q}$ factor of 5.76~million is obtained from a resonance with mode splitting (Fig.~\ref{fig:design}(e)), where we adopted the widely used model for fitting splitting resonances (see the Supporting Information for more details) \cite{Herr2014ModeMicroresonators, Li2024SiliconCommunications, Wei2013ThermalCrystals}. Such mode-splitting is originated from the lifted frequency degeneracy by the clockwise and counter-clockwise modes propagating in opposite directions. It is caused by the interfacial inhomogeneities, e.g. sidewall roughness by fabrication imperfection here. We also characterize 16.5-\textmu m-radius microring resonators and obtain similar ${\rm Q}$ performance (see the Supporting Information for more details). It is noted that the realized ${\rm Q}$ factor is on par with the state-of-art SiC microring resonator \cite{Guidry2022QuantumMicrocombs} but with a more than two times smaller effective area.

Due to its normal GVD, the device cannot support bright solitons. The measured ${\rm Q}$ factor values of microring resonators with different waveguide dimensions are shown in Fig.~\ref{fig:design}(f). The ${\rm Q}$ factor increases when the waveguide width is enlarged, inferring that the light scattering at the sidewall dominates the microring resonators' linear loss. We also note that the existence of a slab compromises the ${\rm Q}$, and the bending loss may cause the reduced ${\rm Q}$ (see the Supporting Information for more details). Therefore, one can choose a waveguide dimension with a small slab thickness for a microring resonator to have a high ${\rm Q}$, a small effective mode area, and anomalous dispersion for efficient octave-spanning comb generation.

\subsection{Octave-spanning Kerr comb generation}

We explore the octave-spanning Kerr soliton generation in 1-THz-FSR microring resonators. Figure~\ref{fig:engineering}(a) shows the SEM image of the microring resonator with a 16.5-\textmu m radius, and the inset shows the cross-section of the waveguide. We characterize the ${\rm Q}$ and GVD of microring resonators with different slab thicknesses (from 50~nm to 100~nm), which are supposed to support the anomalous GVD and DW generation. Fig.~\ref{fig:engineering}(b) shows the histogram of the intrinsic linewidth of microring resonators with a 100-nm thick slab, indicating a most probable value of 80~MHz, corresponding to an intrinsic ${\rm Q}$ factor of 2.4 million. Figure~\ref{fig:engineering}(c) shows the simulated integrated dispersion (${\rm D_{int}}$) of microring resonators with different slab thicknesses, where ${\rm D_{int} = \omega_{\mu} - \omega_0 - D_1\mu = \sum_{n \ge 2} {D_n \mu^n}/{n!}}$ with ${\rm D_n}$ representing the ${\rm n}$th-order dispersion coefficient. As can be seen, for ${\rm h = 0}$~nm, GVD is normal. The anomalous dispersion with two DWs can be achieved with the slab thickness (${\rm h}$) varying from 100~nm to 200~nm. Figure~\ref{fig:engineering}(d) shows measured ${\rm D_{int}}$, where the frequency axis is calibrated through a free space cavity with FSR around 100~MHz \cite{Zheng20194H-SiCPhotonics}. By fitting the data, second-order dispersion (${\rm D_2/2\pi}$) for these four dimensions is extracted as $-$171.9, 51.8, 70.2 and 86.7~MHz, corresponding to GVD values of $-$206, 15, 39, and 48~ps/nm/km at 1560~nm. It matches well with calculated GVD of $-$182, 25, 50 and 58~ps/nm/km.

We pump the device (${\rm W}$, ${\rm H}$ and ${\rm h}$ of 950~nm, 340~nm and 100~nm) with a continuous-wave (CW) laser. Optical parametric oscillation (OPO) threshold power is measured for resonances with different coupling conditions. During the measurement, a pump laser is amplified in an erbium-doped fiber amplifier (EDFA) and coupled into the ${\rm TE_{00}}$ mode of the waveguide through edge coupling. As shown in Fig.~\ref{fig:engineering}(e), the achieved lowest threshold on-chip power for OPO is measured to be 0.29~mW. The dashed line corresponds to the critical-coupling condition. Detailed information can be read in the Supporting Information.

As mentioned in the introduction, the thermal effect is critical for soliton generation. Since SiC provides high thermal conductivity ($\sim$390~$\text{W m}^{-1}\text{K}^{-1}$) \cite{Wei2013ThermalCrystals}, we explore the influence of a slab layer on the thermal shift of the microring resonator resonances. Figure~\ref{fig:engineering}(f) shows the measured normalized transmission with different input power for the microring resonators with a slab thickness of ${\rm h = 0}$~nm and ${\rm h = 100}$~nm. With coupled power of about 3~mW, the resonance shift is around 44~pm and 26~pm for SiC microring resonators with ${\rm h}$ of 0~nm and 100~nm. It should be noted that both microring resonators exhibit similar intrinsic ${\rm Q}$s ($\sim$1.8~million) and coupling conditions to ensure the same intracavity power. 

As shown in Fig.~\ref{fig:engineering}(g), the measured resonance shifts scale linearly with the coupled power. The linear fitting of the measured data shows a 33\% reduction of resonance shift for the device with a SiC slab layer. As both the thermal and Kerr effects contribute to the resonance shift, the thermally-induced resonance shift reduction is more than 33\%, indicating that the thin slab layer improves heat dissipation and is thus beneficial to soliton generation.

With optimized dispersion and thermal properties of the device, we increase the pump power to generate the Kerr comb. Octave-span Kerr combs are successfully generated with 80-mW on-chip power for devices with different ${\rm h}$. However, all generated combs are in a noisy state, and we didn't observe any soliton generation by fast-sweeping the single pumping. The observed resonance shift (100~GHz) at this operation power indicates a considerable thermal effect.

\begin{figure*}
\centering
\includegraphics[width=15.5cm]{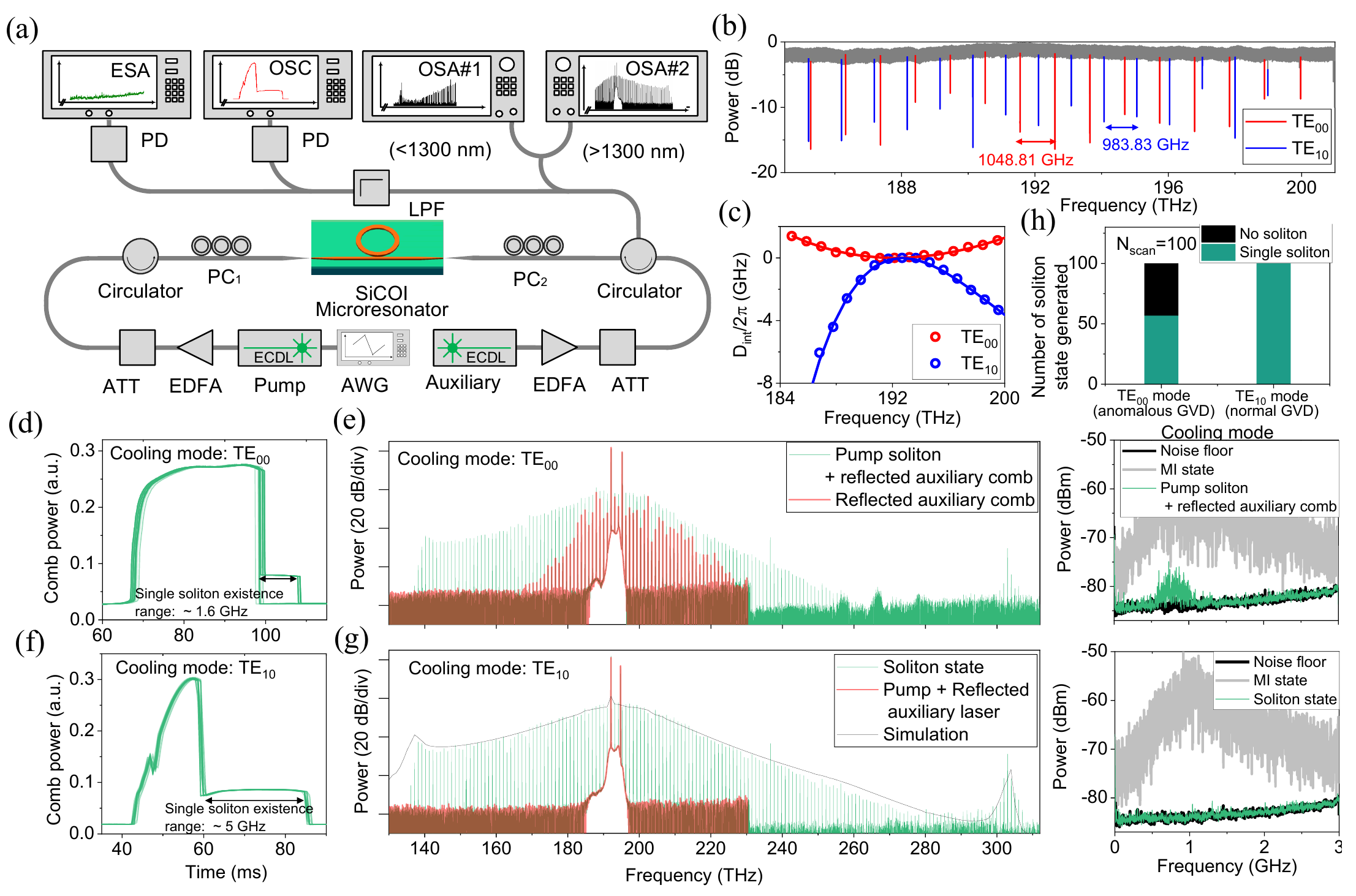}
\caption{\label{fig:octave} \textbf{Deterministic generation of octave-spanning single soliton comb.} (a) Experimental setup for the deterministic soliton generation demonstration. AWG, arbitrary waveform generator. ECDL, external cavity diode laser; ATT, attenuator; PC, polarization rotator; EDFA, erbium-doped fiber amplifier; LPF, long pass filter; PD, photodetector; OSA, optical spectrum analyzer; OSC, oscilloscope. ESA, electronic spectrum analyzer. (b) Transmission spectrum of a multi-mode microring resonator with the waveguide dimension (${\rm H = 340}$~nm, ${\rm W = 950}$~nm, ${\rm h = 100}$~nm, and ${\rm \theta = 75^{\circ}}$). (c) Measured integrated dispersion of the microring resonator for the ${\rm TE_{00}}$ mode and the ${\rm TE_{10}}$ mode. (d, f) Measured output comb power with 100 consecutive pump detuning sweeps using the ${\rm TE_{00}}$ mode (anomalous dispersion) (d) and the ${\rm TE_{10}}$ mode (normal dispersion) (f) as the laser cooling modes. The same pump and auxiliary laser power are used to characterize the same device. (e, g) Measured comb spectra and corresponding low-frequency noise characteristics for the comb generation in (d) and (f), respectively. The black line in (g) is the simulated single soliton comb spectrum envelope. (h) Statistics of soliton generation using ${\rm TE_{00}}$ and ${\rm TE_{10}}$ as the laser cooling mode. The coupled pump power and the cooling laser power are kept at 60~mW and 50~mW, respectively, for the comparison experiments.}
\end{figure*}

\begin{figure*}
\centering
\includegraphics[width=15.5cm]{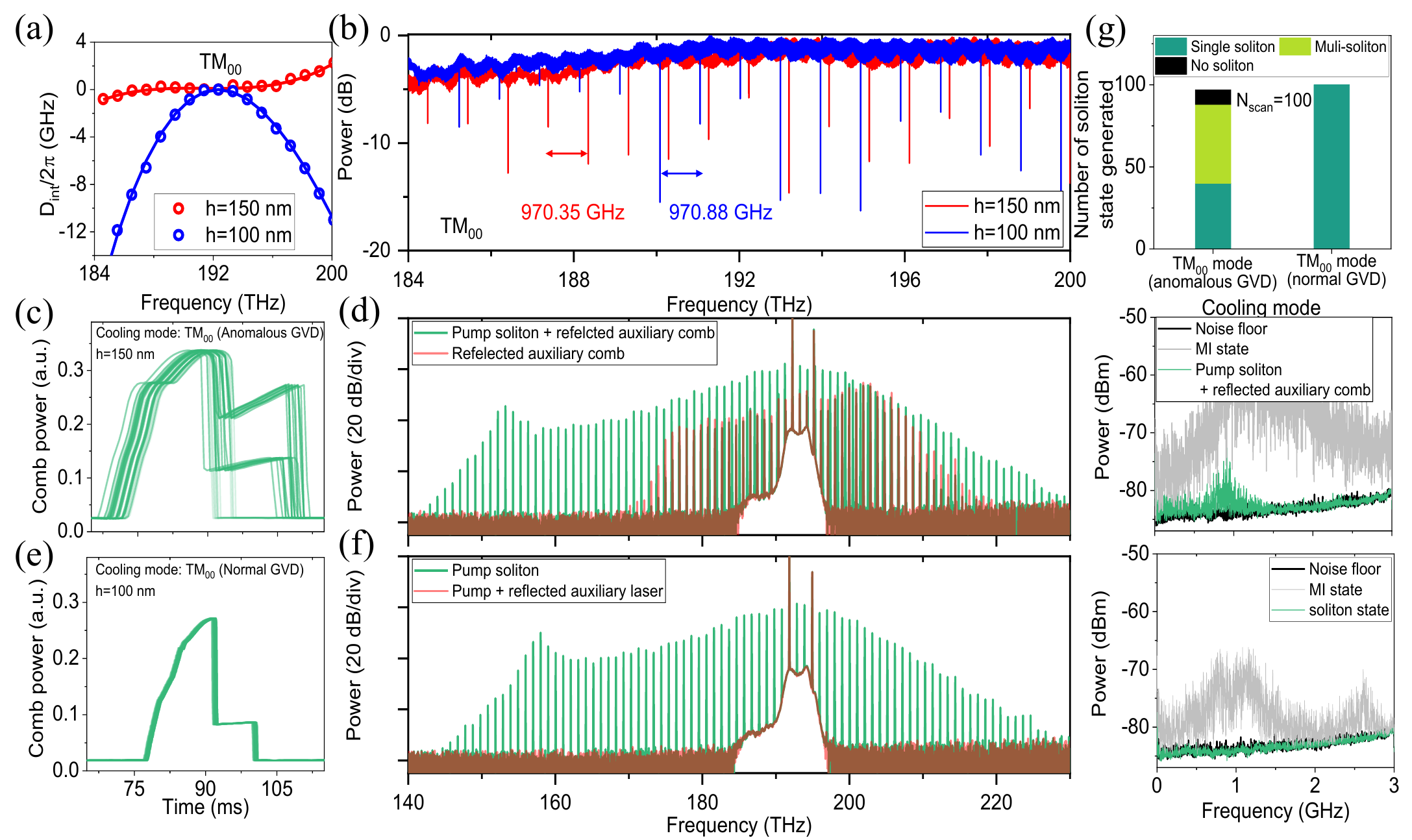}
\caption{\label{fig:TMcooling} \textbf{Comparison of soliton generation using the ${\rm TM_{00}}$ mode as the laser cooling mode with different dispersion properties.} Measured integrated dispersion (a) and transmission spectra (b) of microring resonators with different slab thicknesses (${\rm h = 100}$~nm, ${\rm h = 150}$~nm) for the ${\rm TM_{00}}$ mode. (c, e) Measured output comb power with 100 consecutive pump detuning sweeps using the anomalous-dispersion ${\rm TM_{00}}$ mode (c) and the normal-dispersion ${\rm TM_{00}}$ mode (e) as the laser cooling mode. The coupled laser power is the same in the laser cooling mode. (d, f) Measured comb spectra and corresponding low-frequency noise characteristics for the comb generation in (c) and (e), respectively. (g) Statistics of soliton generation using ${\rm TM_{00}}$ mode with anomalous and normal ${\rm GVD}$ as the laser cooling mode. The coupled pump power and the cooling laser power are kept at 63~mW and 80~mW, respectively, for the comparison experiments.}
\end{figure*}

\subsection{Deterministic generation of octave-spanning soliton comb}

Here, we investigate the soliton generation condition by using the laser cooling technique and compare the soliton characteristics concerning their spectrum bandwidth, noise characteristics, and the success rate of single soliton generation when different spatial modes are selected as the laser cooling mode. 

The experimental setup is shown in Fig.~\ref{fig:octave}(a) and a detailed description can be found in the Supporting Information. Figure~\ref{fig:octave}(b) shows the measured transmission spectrum of a microring resonator that supports octave-spanning comb generation. The waveguide width, height, and slab thickness are 950~nm, 340~nm, and 100~nm, respectively. The ${\rm TE_{00}}$ and ${\rm TE_{10}}$ modes are marked as red and blue in Fig.~\ref{fig:octave}(b). The FSR of these two modes is 1048.81~GHz and 983.83~GHz. As shown in Fig.~\ref{fig:octave}(c), the measured GVD of the ${\rm TE_{00}}$ and ${\rm TE_{10}}$ modes are 15~ps/nm/km and $-$378~ps/nm/km. The anomalous dispersion of the ${\rm TE_{00}}$ mode allows for KCG while the normal dispersion of the ${\rm TE_{10}}$ mode prohibits KCG. It is noted that the mean intrinsic ${\rm Q}$ of ${\rm TE_{10}}$ mode is 2~million (see the Supporting Information for more details). We compare the soliton characteristics by using either the ${\rm TE_{00}}$ mode or the ${\rm TE_{10}}$ mode as the laser cooling mode. In both cases, the ${\rm TE_{00}}$ mode is used as the comb operation mode. First, we use the ${\rm TE_{00}}$ mode as the laser cooling mode. During the measurement, the auxiliary laser is tuned into a resonance at 1536~nm and the pump laser scans over a 15-GHz frequency range across another resonance at 1561~nm with a tuning speed of 200~GHz/s. A long pass filter is applied to filter spectra with a wavelength longer than the pump for monitoring the comb power. The power trace is recorded in the oscilloscope (OSC) during the pump tuning. Figure~\ref{fig:octave}(d) shows 100 consecutive experimental traces overlaid for the monitored comb power. The observed step corresponds to the single soliton state. It is noted that the obtained longest SER of 1.6~GHz is realized through suppressing the thermal effect to the maximum extent by optimizing the auxiliary laser frequency. However, the single soliton step is absent in 43 traces. To study the soliton characteristic, we analyze the spectrum of generated comb and its low-frequency noise shown in Fig.~\ref{fig:octave}(e). Surprisingly, we observe noise at the output. Further investigation of the optical spectrum shows that a MI comb (red in the optical spectrum) exists when the pump laser is tuned out of the resonance. As the soliton comb and the reflected MI comb spectra overlap in spectrum since the same mode is used for the comb operation and laser cooling mode, they cannot be distinguished in the output spectrum. The noise can be attributed to the counter-propagating MI comb produced by the auxiliary laser, which can be either reflected by the imperfect sidewall of the microring resonator or the sample facet. Our measurements indicate that the facet reflection exceeds 1\%. Additionally, the MI comb might lead to cross-phase modulation or interference with the soliton comb, giving rise to extra noise \cite{Lucas2018SpatialMicrocombs}.

It is essential to maintain thermal equilibrium as much as possible in the microring resonator when the pump laser is tuned across the resonance for deterministic single soliton generation \cite{Zhao2021SolitonLaser}. This requires a minimal change of the resonance shift. The resonance shift (${\rm \Delta \omega_{total}}$) is determined by the ${\rm Q}$-power product of the total intracavity field, ${\rm \sum Q_i P_i}$, where ${\rm Q_i}$ and ${\rm P_i}$ are the quality factor and intracavity power of the pump and auxiliary laser \cite{Zhao2021SolitonLaser}. The ideal condition corresponds to the unchanged ${\rm \sum Q_i P_i}$ during the pump tuning. It can be qualitatively understood that as the pump laser is tuning from the blue-detuned side to the red-detuned side of a resonance while the auxiliary laser wavelength is fixed at the blue-detuned side of the resonance, the ${\rm P_p}$ and ${\rm P_a}$ keep increasing and decreasing, respectively, to maintain a minimal change of ${\rm \sum Q_i P_i}$. When the pump laser is tuned across the resonance and enters into the red-detuned region, there is an abrupt reduction of ${\rm P_p}$ results in blue shift of the resonance. However, the resonance shift can be effectively compensated if the auxiliary laser is properly adjusted because the blue-shifted resonance will heat up the resonance by an increase of ${\rm P_a}$. Since part of the power is used for MI comb generation when ${\rm TE_{00}}$ is used for cooling, the auxiliary laser is not efficiently used for cooling. Therefore, the amount of power used for cooling varies in each scan, which leads to the failure of deterministic octave-span microcomb generation. 

To efficiently use the auxiliary laser for cooling, the comb generation in the laser cooling mode needs to be prohibited. As the ${\rm TE_{10}}$ mode is designed to have normal dispersion, the same experiment is performed when we use the ${\rm TE_{10}}$ mode as the laser cooling mode. In the experiment, the on-chip pump and auxiliary laser power are set to be 60~mW and 50~mW, respectively. The auxiliary laser is carefully adjusted to elongate the SER and we obtained an SER of 5~GHz as shown in Fig.~\ref{fig:octave}(f), which is orders of magnitude larger than the previously realized soliton \cite{Guidry2022QuantumMicrocombs, Wang2022SolitonPlatform}. One can find that all 100 overlaid consecutive comb power traces show the soliton step with the same SER, inferring deterministic soliton generation. The generated soliton can be stably maintained for several hours in the laboratory environment without any active feedback. Further increase of cooling laser power helps to extend the SER because less thermal resonance shift will be caused by the pump laser \cite{Nishimoto2022ThermalSideband}. Figure~\ref{fig:octave}(g) shows the spectrum of the generated soliton (green) spans over one octave from 136 to 307~THz with two dispersive waves at 138~THz ($\sim$990~nm) and 302~THz ($\sim$2171~nm), respectively. The reflected spectrum (red curve in Fig.~\ref{fig:octave}(g)) confirms that the comb generation is prohibited in ${\rm TE_{10}}$ mode. The simulated spectrum (black line) matches well with the measured spectrum. The spectral envelope verifies that only one soliton for ${\rm TE_{00}}$ mode is generated. Fig.~\ref{fig:octave}(h) summarizes the statistical result. The success rate of accessing soliton is 57\% and 100\% for the laser cooling mode being the ${\rm TE_{00}}$ mode and the ${\rm TE_{10}}$ mode, respectively. Therefore, normal-dispersion laser cooling is a robust method for the deterministic generation of single soliton. It is noted that comparable intracavity power of the pump laser and cooling laser is favored in the experiment for single deterministic soliton generation. However, critical coupling condition is for both pump and cooling modes beneficial to lower the required power.

The proposed method can be versatile since microring resonators can support different spatial modes and their dispersion can be engineered for the laser cooling purpose. To demonstrate this, we now select the ${\rm TM_{00}}$ mode as the laser cooling mode and investigate the soliton generation. We characterized two devices that have the same waveguide width (750~nm) and waveguide height (340~nm) but different slab thicknesses of 100~nm and 150~nm. The GVD of the comb operation modes (${\rm TE}$) in these two devices have similar values (53~ps/nm/km and 58~ps/nm/km). While the GVD of the laser cooling modes (${\rm TM_{00}}$) have different signs as shown in Fig.~\ref{fig:TMcooling}(a) (2~ps/nm/km and $-$423~ps/nm/km). Figure~\ref{fig:TMcooling}(b) is the transmission spectra of ${\rm TM_{00}}$ mode in resonators with ${\rm h = 150}$~nm (red) and ${\rm h = 100}$~nm (blue). We use the same experimental setup and pumping procedures to compare the comb characteristics when ${\rm TM_{00}}$ modes with opposite signs of GVD are used as the laser cooling modes. Figure~\ref{fig:TMcooling}(c) shows 100 consecutive comb power traces using anomalous ${\rm TM_{00}}$ mode for cooling. The statistics indicate that except for single soliton generation, it can either be in a multi-soliton state or no soliton. The result in Fig.~\ref{fig:TMcooling}(g) shows the occurrence rate for single-, multi-soliton and no-soliton states are 40\%, 48\%, and 12\%, respectively. As shown in Fig.~\ref{fig:TMcooling}(d), two combs can be observed in the optical spectrum, where the soliton (green) and MI comb (red) are generated in the ${\rm TE_{00}}$ mode and ${\rm TM_{00}}$ mode, respectively. Similar noise characteristics are obtained as in Fig.~\ref{fig:TMcooling}(e). It confirms that the reflected MI comb generated by the laser cooling mode adds noise to the output soliton. In comparison, normal ${\rm TM_{00}}$ mode as the laser cooling mode is applied for soliton generation. Fig.~\ref{fig:TMcooling}(e, f) shows the comb generation characteristics and it turns out that utilizing the normal-dispersion ${\rm TM_{00}}$ mode for cooling enables single soliton generation with a success rate of 100\% as shown in Fig.~\ref{fig:TMcooling}(g). Moreover, a clean soliton spectrum is obtained (Fig.~\ref{fig:TMcooling}(f)) as well as the low-noise characteristics (Fig.~\ref{fig:TMcooling}(h)). The results confirm that normal-dispersion laser cooling is a versatile method for deterministic soliton generation. To test the viability of the proposed technique as a universal applicable approach across different material platforms, we have verified its applicability in the widely recognized ${\rm Si_3N_4}$ platform (see the Supporting Information for more details). In this platform, we have successfully demonstrated a significant enhancement in the success rate of single soliton operation.

\section{\label{sec:conclusion}Discussion and Conclusion}

Though we achieve an octave-spanning soliton comb, the relatively low power level at two dispersive wave frequencies could make it challenging for f-2f self-referencing. It is partly because of the finite bandwidth of the ring-to-bus coupler. A possible solution is to design a broadband directional coupler to efficiently couple the light out at dispersive wave frequencies \cite{Moille2019Kerr-MicroresonatorTemperatures, Rao2021TowardsCombs}. Moreover, advanced dispersion engineering could also assist to achieve low and flat anomalous dispersion for more efficient energy transfer to dispersive waves \cite{Lucas2022TailoringMicroresonators}. The required power for both the pump and auxiliary lasers for the demonstrated octave-spanning single soliton is compatible with on-chip DFB lasers \cite{Mashanovitch2018High-PowerLinks}. The operation power can be further reduced by improving the ${\rm Q}$ and optimizing the dispersion design for fully etched microring resonators. For other applications that require narrow bandwidth like telecommunication \cite{Riemensberger2020MassivelyMicrocomb, Hu2018Single-sourceTransmission}, less soliton operation power (see the Supporting Information for more details) also eases the power requirement of on-chip laser for heterogeneous integration.

In conclusion, we experimentally demonstrate the first low-power operated octave-spanning soliton microcomb via dispersive wave generation in the 4H-SiCOI platform. The achieved high ${\rm Q}$, high confinement SiCOI microring resonators greatly reduce the optical parametric oscillation threshold power to the sub-milliwatt level. A deterministic single soliton generation method is proposed and verified by utilizing the normal-dispersion laser cooling technique. The proposed normal-dispersion laser cooling requires dispersion management of multi-mode microresonators. Material platforms with large thermo-optic coefficients can benefit from this method for deterministic soliton generation. Compared with other material platforms where octave-spanning comb generation has been demonstrated, SiC material provides decent intrinsic quadratic nonlinearity \cite{Song2019Ultrahigh-QCarbide, Lukin20204H-silicon-carbide-on-insulatorPhotonics, Zheng2025EfficientNanowaveguides}. It could ease the fabrication complexity when both octave-spanning comb generation and frequency doubling are required on the same chip (see the Supporting Information for more details). Our results provide great potential for chip-scale octave-spanning soliton microcomb and present a step towards applications like integrated optical atomic clocks and frequency synthesizers.\\

\textbf{Availability of data and materials}\\
The datasets generated and/or analyzed during in this study are available from the corresponding authors upon reasonable request.\\

\textbf{Competing interests}\\
The authors declare no competing financial interests.\\

\textbf{Funding}\\
This work is supported by European Research Council under the EU's Horizon 2020 research and innovation programme (grant agreement no. 853522, REFOCUS), Independent Research Fund Denmark (ifGREEN 3164-00307A), Danish National Research Foundation (SPOC ref. DNRF123), Innovationsfonden (Green-COM 2079-00040B), National Key R\&D Program of China (Grant No. 2022YFA1404601, No. 2022YFA1404602), the Chinese Academy of Sciences Project for Young Scientists in Basic Research (Grant No. YSBR-69), and the National Natural Science Foundation of China (Grant Nos. 62293521, 12575313, 12074400, and 62205363).\\


 \bibliography{references}


\end{document}